\documentclass[12pt]{article}
\usepackage{amssymb}
\title{\Large \bf Note on a novel vortex dynamics of spacetime as a heuristic model of the vacuum energy}
\author{\small Hirofumi Sakuma\\
\small Yokohama Institute for Earth Sciences, JAMSTEC \\
\small e-mail: sakuma@jamstec.go.jp}
\date{}
\begin{document}
\maketitle
\begin{abstract}
Vortex or spin is an important and ubiquitous form of motions existing in almost all scale ranges of the universe and its dynamics 
is still an active research theme in the classical as well as modern
physics. As a novel attempt of such studies, here we show that a class of vortex dynamics generated by newly defined Clebsch parametrised (CP) flows parallel to geodesics exhibits an intriguing property that it is isomorphic to the spacetime structure itself on which it is defined in the sense that its energy-momentum conservation equation automatically assumes exactly the same form as the Einstein field equation. Implications of the existence of such a model is briefly discussed from the view point of a current hot cosmological interest on dark energy together with elusive concept on gravitational energy radiation.
\end{abstract}

\section{\large Introductory remarks and fixing conventions}

Since a vortex model we are going to discuss here is not for describing vortical modes of some material medium,
but for the intrinsic properties of spacetime, it is quite natural to start our discussion from 
electromagnetic (EM) radiation field which played a leading role in constructing special theory of relativity as the physics of a flat spacetime.
The mathematical relation between  skew-symmetric EM field $F_{\mu\nu}$ and EM vector potential $A_{\mu}$
is exactly the same as that of vortex tensor and its associated velocity vector in hydrodynamics, but a crucial physical difference
would be that $A_{\mu}$ is a potential that cannot be determined completely. Concerning this point, the question
whether $A_{\mu}$ is unphysical or not in contrast to $F_{\mu\nu}$ has long
been the main target of the debates involving the interpretation of the
Bohm-Aharonov (AB, for short) effect, and the standard understanding of the present situation is 
that physical relevance of the vector potential $A_{\mu}$ has been established by the clear-cut experiments performed by Tonomura et. al. \cite{RefJ1}. However, these experiments cannot be taken as the evidence for the tangibility (\it i.e.\rm,  complete freedom) of all the four components of $A_{\mu}$, 
 since what is actually relevant to the AB effect is not $A_{\mu}$ itself but its spatial loop integral which is
related to the rotational part of $A_{\mu}$ through Stokes' theorem.
The arguments on AB effect does not claim anything about the gauge dependent irrotational part of
$A_{\mu}$. 
It is considered to be unphysical from a conventional viewpoint of gauge theory, though
there seems to be a subtle issue concerning the problem of observability
expressed in terms of existing probability which is missing in the classical physics but it becomes a central concern in quantum physics. As an example illuminating this point, in section 5, we consider   
 Nakanishi-Lautrup (NL) formalism \cite{RefJ2} on manifestly covariant quantisation of EM field, for
which it can be shown by  Ojima \cite{RefJ3} that the gauge-fixing (GF) part plays certain physical roles \it macroscopically.
\rm  Of course, detailed consideration on this subtle issue covering quantum physics is beyond the scope of the present discussion, however, based on the energy-momentum conservation, we will point out a hitherto unreported possibility that B field given in NL formalism may have physical relevance at least classically.

The fact that $A_{\mu}$ has a certain physical relevance, albeit it may carry the features of potential 
quantity in the sense mentioned above, motivates us to reexamine the EM (radiation) theory from the hydrodynamic viewpoint. 
As a starting point of such an attempt, let us take a view that free EM radiation field is regarded as a vortex dynamics of null  
geodesics.  Combining this view with the above-mentioned physical relevance of $A_{\mu}$ leads us to 
study a possibility of hydrodynamic model in which $A_{\mu}$ is restricted to move along a certain
null geodesic just like velocity vector $v_{\mu}$ of an electrically neutral point mass moving under the
influence of gravitational field.
In this respect, we can say that our approach shares a background similar to the twistor theory 
in terms of complex numbers, since, as is shown shortly,
light-like modes of our vortical system is closely related to the notion of null geodesics called shear-free null congruence
studied by Robinson \cite{RefJ4} which corresponds, in the case of EM field, to a null vector 
parallel to $4d$ Poynting vector. 

Now we start with fixing conventions and defining several useful terms. As the
terminology of the spin dynamics developed in twistor theory is useful, we
follow some conventions employed by Penrose and Rindler \cite{RefB1} of which
brief introduction is given, say, by Huggett and Tod \cite{RefB2}. By
$\eta_{\mu\nu}$ and $g_{\mu\nu}(0\leq\mu,\nu\leq3)$, we denote, respectively,
the Minkowski metric $diag(1,-1,-1,-1)$ in an orthonormal frame and the
Lorentzian metric tensor for a pseudo Riemannian manifold $M$. A covariant
vector $U_{\mu}$ with a lower index $\mu$ is simply called a covector.
A skew symmetric second-rank tensor $X_{\mu\nu}=-X_{\nu\mu}$ is referred
here to as a bivector $(X_{01},X_{02},X_{03};X_{32},X_{13},X_{21})$: for
instance, for $X_{\mu\nu}=F_{\mu\nu}=\partial_{\mu}A_{\nu}-\partial_{\nu}A_{\mu}$: electromagnetic field strength, we have
\begin{equation}
F_{\mu\nu}=\left(
\begin{array}
[c]{cccc}
0 & E^{1} & E^{2} & E^{3}\\
-E^{1} & 0 & -B^{3} & B^{2}\\  
-E^{2} & B^{3} & 0 & -B^{1}\\
-E^{3} & -B^{2} & B^{1} & 0
\end{array}
\right)  =(\vec{E},\vec{B}). \label{eqn:2000}
\end{equation}
A bivector $X_{\mu\nu}$ is said to be simple if it can be written as the
exterior product of two vectors $2U_{[\mu}V_{\nu]}$ (or, $X=\frac{1}{2}
X_{\mu\nu}dx^{\mu}\wedge dx^{\nu}=U\wedge V$) where the square bracket denotes
anti-symmetrisation defined by 
\begin{equation}
X_{[\mu...\nu]}=\frac{1}{r!}\sum_{\sigma\in\mathfrak{S}_{r}}(sgn(\sigma))X_{\sigma(\mu)...\sigma(\nu)}
\label{eqn:1001}
\end{equation}
for $X${$_{\mu...\nu}$ having $r$ indices and the sum taken over all
permutations $\sigma$}$\in\mathfrak{S}_{r}$ {with signatures $sgn(\sigma)$.
The totally anti-symmetric tensor $\epsilon_{\mu\nu\rho\sigma}$ is defined by
$\epsilon_{\mu\nu\rho\sigma}=\epsilon_{\lbrack\mu\nu\rho\sigma]},$%
{$\epsilon_{0123}=1$} and $\epsilon_{\mu\nu\rho\sigma}\epsilon^{\mu\nu
\rho\sigma}=-24$. With the aid of $\epsilon_{\mu\nu\rho\sigma}$, $\eta_{\mu
\nu}$ and $\eta^{\mu\nu}$, we can define the Hodge dual $(^{\ast}F)_{\mu\nu}$
of a bivector $F_{\mu\nu}$ by
\begin{eqnarray}
(^{\ast}F)_{\mu\nu} &  =-\frac{1}{2}\epsilon_{\mu\nu}^{\ \ \rho\sigma
}F_{\rho\sigma};\,\, \label{eqn:2201} \\
(^{\ast}F)^{\mu\nu} &  =-\frac{1}{2}\epsilon_{\ \ \rho\sigma}^{\mu\nu
}F^{\rho\sigma};\,\, \label{eqn:2202} \\
\textnormal{e.g., }(^{\ast}F)_{01} &  =F_{32},\,\,\,\,(^{\ast}F)_{32}=-F_{01}. \label{eqn:2203} 
\end{eqnarray}
In terms of the above notation, we
consider the field of geodesics on a given $4d$ pseudo Riemannian manifold $M$.
The equation describing a geodesic assumes the form of 
\begin{equation}
(\nabla_{U}U)_{\mu}=U^{\nu}\nabla_{\nu}U_{\mu}=U^{\nu}(\nabla_{\nu}U_{\mu
}-\nabla_{\mu}U_{\nu})+\nabla_{\mu}(U^{\nu}U_{\nu}/2)=0  \label{eqn:1003}
\end{equation}
where $\nabla_{X}$ denotes the covariant derivative along a vector field
$X=X^{\mu}\partial_{\mu}$ associated with the Levi-Civita connection.
In case that $U^{\mu}$ denotes velocity four vector, that is to say, it can
be defined as $U^{\mu}=dx^{\mu}/ds$ where
$x^{\mu}$ are Lagrangian coordinates of a given (fluid) particle and $ds^{2}=g_{\mu\nu}dx^{\mu}dx^{\nu}$ is the 
infinitesimal distance measured on the trajectory of the particle, the magnitude
of it is respectively normalised as
\begin{eqnarray}
g_{\mu\nu}U^{\mu}U^{\nu}=U_{\nu}U^{\nu}=0;\;\;\;\;\; 
g_{\mu\nu}U^{\mu}U^{\nu}=U_{\nu}U^{\nu}=\pm1,   \label{eqn:1005}
\end{eqnarray}
depending on whether $U^{\mu}$ is null or not. 
In what follows, we are going to discuss light-like modes first mainly because it seems to be related to
a problem of gravitational energy radiation: another point in dispute in addition to dark energy
issue, where new notions on the classical wave-particle duality and on synchronised energy-carrying 
gravito-EM modes are discussed based on the possible macroscopic physical relevance of B field.
 Non-light-like modes together with the central result of this note are given in the final section 6.

\section{Vortex dynamics of light-like CP flows}
\label{sec:1}

The key representation for $U_{\mu}$ employed in our note is a newly introduced parametrisation  
we tentatively call modified CP defined
by
\begin{equation}
U_{\mu}=\frac{1}{2}(\lambda\nabla_{\mu}\phi-\phi\nabla_{\mu}\lambda),  \label{eqn:1006}
\end{equation}
where $\lambda$ and $\phi$ are a couple of complex scalar fields. 
CP flow\cite{RefB3} has been extensively studied in the field of Hamiltonian formulation
 of barotropic fluid, where fluid velocity $v_{\mu}$ is basically parametrised with three scalars:
\begin{equation}
v_{\mu}=\nabla_{\mu}\chi + \lambda\nabla_{\mu}\phi.  \label{eqn:1007}
\end{equation}
The reason why a new parametrisation shares the name of CP is because both parametrisations
essentially depend on the characteristic form of $\lambda\nabla_{\mu}\phi$, though 
the irrotational component $\nabla_{\mu}\chi$ in the conventional CP is missing in the modified one where a couple of variables $\lambda$ and $\phi$ are treated on equal footing.  
Before looking into the consequences of (\ref{eqn:1006}) imposed on $U_{\mu}$ of 
non-light-like modes, we first consider the case which corresponds to free propagation of EM waves.
A class of solutions of particular importance for such a case is the null
geodesics with wave property, namely, the solutions to d'Alembert equation for
a massless field $B$:
\begin{equation}
g^{\mu\nu}\nabla_{\mu}\nabla_{\nu}B=0;\,\,\,\,g^{\mu\nu}\nabla_{\mu}
B\nabla_{\nu}B=0. \label{eqn:1008}
\end{equation}
A couple of equations given in (\ref{eqn:1008}) are well-documented ones whose compatibility
is readily checked for a plane wave of the form: $B=exp(ik_{\nu
}x^{\nu})$ with $k^{\nu}k_{\nu}=0$ in a flat spacetime. (\ref{eqn:1008}) assumes that
the the solutions to d'Alembert equation for a massless field exist on a curved spacetime keeping
the compatibility conditions unchanged. With this $B$, for a light-like vector $U_{\mu}$,
 we can introduce CP flow as
\begin{equation}
U_{\mu}=\lambda\nabla_{\mu}B,  \label{eqn:1009}
\end{equation}
where $\lambda$ is another Clebsch variable to be determined.
The fact that the expression (\ref{eqn:1009}) can be a special case of (\ref{eqn:1006}) for light-like
modes will be touched on in the final section 6 where we deal with non-light-like modes. Although the
complex nature of Clebsch variables are fully exploited there, for simplicity, we assume that all the variables
are real in the following analyses on light-like modes.

For notational simplicity, let us define covectors $L_{\mu}$, $C_{\mu}$ and a simple bivector $S_{\mu
\nu}$ constructed by them as
\begin{equation}
L_{\mu}\equiv\nabla_{\mu}\lambda;\,\,\,\,C_{\mu}\equiv\nabla_{\mu
}B;\,\,\,\,S_{\mu\nu}\equiv\nabla_{\nu}U_{\mu}-\nabla_{\mu}U_{\nu}=C_{\mu
}L_{\nu}-L_{\mu}C_{\nu}. \label{eqn:1012}
\end{equation}
Substituting (\ref{eqn:1009}) into the left-hand side of (\ref{eqn:1003})
with the use of self-orthogonality condition in (\ref{eqn:1008}), we have
\begin{equation}
U^{\nu}\nabla_{\nu}U_{\mu}=S_{\mu\nu}(\lambda C^{\nu})=(C_{\mu}L_{\nu}-L_{\mu}
C_{\nu})(\lambda C^{\nu})=(L_{\nu}C^{\nu})\lambda C_{\mu}.  \label{eqn:2105}
\end{equation}
So, if we specify $L_{\mu}$ such that
\begin{equation}
C^{\nu}\nabla_{\nu}L_{\mu}=0,    \label{eqn:2001}
\end{equation}
then, it follows immediately that we have  
\begin{eqnarray}
L^{\mu}(C^{\nu}\nabla_{\nu}L_{\mu})=0;\;\;\;\;& \Rightarrow &\;\;\; 
   C^{\nu}\nabla_{\nu}(L^{\mu}L_{\mu})=0;  \label{eqn:2003}  \\
C^{\mu}(C^{\nu}\nabla_{\nu}L_{\mu})=0;\;\;\;\;& \Rightarrow &\;\;\;
   C^{\nu}[\nabla_{\nu}(C^{\mu}L_{\mu}) -L_{\mu}\nabla_{\nu}C^{\mu}] \nonumber \\
 & = & C^{\nu}\nabla_{\nu}(C^{\mu}L_{\mu})=0.  \label{eqn:2005} 
\end{eqnarray}
In deriving (\ref{eqn:2005}), the use has been made of $C^{\nu}\nabla_{\nu}C^{\mu}=0$,
which says that $C^{\nu}$ itself satisfies the null geodesic equation. Note that  
(\ref{eqn:2005}) allows us to have the following orthogonality condition between
the two vectors $L^{\mu}$ and $C^{\mu}$.
\begin{equation}
L_{\nu}C^{\nu}=0. \label{eqn:1014}
\end{equation}
Since any vector perpendicular to a given null vector is either the same null
vector or a spacelike one, we can choose $L_{\nu}$ such that it is a spacelike
 vector satisfying (\ref{eqn:1014}), namely,
\begin{equation}
\rho \equiv -L^{\nu}L_{\nu}> 0,  \label{eqn:2007}
\end{equation}
of which important implication will be discussed later based on the central result
presented in the final section.
With the orthogonality condition (\ref{eqn:1014}), (\ref{eqn:2105}) becomes
\begin{equation}
U^{\nu}\nabla_{\nu}U_{\mu}=S_{\mu\nu}(\lambda C^{\nu})= S_{\mu\nu}U^{\nu}=0,  \label{eqn:1015}
\end{equation}
which says that CP flow (\ref{eqn:1009}) satisfies a null geodesic equation.
It is worthwhile to point out a similarity between (\ref{eqn:1015}) and the following 
equation for the EM field:
\begin{equation}
F_{\mu\nu}P^{\nu}=0, \label{eqn:1016}
\end{equation}
where $F_{\mu\nu}$ and $P^{\nu}$ denote, respectively, the EM bivector and
Poynting 4-vector. The latter one is parallel to a null geodesic perpendicular
to both the electric $\vec{E}=(F_{01},F_{02},F_{03})$ and the magnetic
$\vec{B}=-(F_{23},F_{31},F_{12})$ bivectors. The difference in form between (\ref{eqn:1015}) and
(\ref{eqn:1016}) is that the bivector $F_{\mu\nu}$ is not defined as the curl of $P_{\mu}$ while
$S_{\mu\nu}$ is derived by the curl of $U_{\mu}$.

One of the direct consequences of (\ref{eqn:1009}) is the following identity:
\begin{equation}
S_{\mu\lbrack\nu}S_{\rho\sigma]}=0 \label{eqn:1017}
\end{equation}
which is directly obtained by substituting the third expression in
(\ref{eqn:1012}) into the left-hand side of (\ref{eqn:1017}) and, from
(\ref{eqn:1017}), we also have
\begin{equation}
D\equiv S_{01}S_{23}+S_{02}S_{31}+S_{03}S_{12}=0, \label{eqn:1018}
\end{equation}
where $D=Pf(S)$ is the Pfaffian of the anti-symmetric matrix $S_{\mu\nu}$:$D^{2}=Det(S_{\mu\nu})$.
Similarly to $\vec{E}\cdot\vec{M}=0$ for the EM field $F_{\mu\nu}$, the orthogonality between
$(S_{01}, S_{02}, S_{03})$ and $(S_{23}, S_{31}, S_{12})$ should hold as a necessary condition for the existence
of non-zero vector $U^{\nu}$ satisfying the matrix equation of (\ref{eqn:1015}) because of $Det(S_{\mu\nu})=D^{2}$.
Then, the matrix elements $S_{\mu\nu}$ have $(6-1)=5$ independent components, and hence, they can be viewed as
the homogeneous coordinates of a projective space with dimensionality $(5-1)=4$.
We recall here, as has been noticed by IO  \cite{RefJ7}, the well-known relations among projective space, Grassmannian manifolds
and Pl\"{u}cker coordinates of the latter: while a Grassmannian manifold $GM(p,n;F)$ defined by
the set of all linear subspaces with a fixed dimensionality $p$ in a given $n$-dimensional linear
space $F^{n}$ over a scalar field $F$ is more general than the concept of a projective space
$PF^{q}=GM(1,q+1;F)$, the Grassmannian manifold can, however, be embedded in a higher dimensional
projective space by means of the so-called Pl\"{u}cker coordinates consisting of minor 
determinants of an anti-symmetric matrix constrained by them.  Adapting this viewpoint to 
the Grassmannian manifold $GM(2;4)$ of all the $2$-dimensional linear subspaces of $R^{4}$, 
we can regard the above orthogonality relation (\ref{eqn:1018}) as the
 Pl\"{u}cker condition among the  Pl\"{u}cker coordinates given by the matrix elements 
$S_{\mu\nu}$. In this context, it is seen that the necessary and sufficient condition for holding
(\ref{eqn:1018}) is given by the simplicity of the bivector $S_{\mu\nu}$
 (see, for instance, Huggett and Tod \cite{RefB2}), which is consistent with  (\ref{eqn:1012}) under the assumption of (\ref{eqn:1009}). \rm

At this point, it would be useful to look into certain geometric properties of
a vector $Q^{\nu}$ perpendicular to a given bivector field $S_{\mu\nu}$, that
is,
\begin{equation}
S_{\mu\nu}Q^{\nu}=0. \label{eqn:1019}
\end{equation}
For later comparison between bivectors $S_{\mu\nu}$ and $F_{\mu\nu}$ of EM
field, it is convenient to introduce a 4-vector $\Pi^{\mu}$, corresponding to
a Poynting vector $P^{\mu}$ in (\ref{eqn:1016}), defined by:
\begin{eqnarray}
\Pi^{0}  &  =S_{S}^{2}\equiv(S_{23})^{2}+(S_{31})^{2}+(S_{12}
)^{2};\,\,\,\,\Pi^{1}\equiv-(S_{02}S_{12}-S_{03}S_{31});\label{eqn:1020}\\
\Pi^{2}  &  \equiv-(S_{03}S_{23}-S_{01}S_{12});\,\,\,\,\Pi^{3}\equiv
-(S_{01}S_{31}-S_{02}S_{23}), \label{eqn:1021}
\end{eqnarray}
where the sign of spatial components of $\Pi^{\mu}$ is chosen identically to
that of the Minkowski metric $diag(1,-1,-1,-1)$ referred to at the
beginning. In Appendix A, we show that the solution $Q^{\nu}$ to (\ref{eqn:1019})
is either a null vector parallel to $\Pi^{\nu}$ or a space-like one.

\section{Vortex dynamics of null geodesics and its dual representations}
\label{sec:2}

The energy-momentum tensor $T_{\mu}^{\nu}$ associated with EM radiation field
is given in its mixed tensor form by 
\begin{equation}
T_{\mu}^{\nu}=-F_{\mu\sigma}F^{\nu\sigma}. \label{eqn:1033}
\end{equation}
The corresponding quantity $\hat{T}_{\mu}^{\nu}$ for $S_{\mu\nu}$ is given by
\begin{eqnarray}
\hat{T}_{\mu}^{\nu} &  =-S_{\mu\sigma}S^{\nu\sigma}=-(C_{\mu}L_{\sigma}
-L_{\mu}C_{\sigma})(C^{\nu}L^{\sigma}-L^{\nu}C^{\sigma})\nonumber\\
&  =-(L_{\sigma}L^{\sigma})C_{\mu}C^{\nu}=\rho C_{\mu}C^{\nu}
,\label{eqn:1034}
\end{eqnarray}
in the use of equations (\ref{eqn:1008}), (\ref{eqn:1014}) and (\ref{eqn:2007}).
While $C^{\mu}$ is light-like, (\ref{eqn:1034}) is identical in form to the
energy-momentum tensor of free moving (fluid) particles.
Actually, 
following from $C^{\nu}\nabla_{\nu}C_{\mu}=0$, $\nabla_{\nu}C^{\nu}=0$ and 
$C^{\nu}\nabla_{\nu}\rho =0$, which are derived by (\ref{eqn:1008}), (\ref{eqn:2003}) and (\ref{eqn:2007}), we get
\begin{equation}
\nabla_{\nu}\hat{T}_{\mu}^{\nu}=0. \label{eqn:1035}
\end{equation}
Since (\ref{eqn:1034}) has dual representations, (\ref{eqn:1035}) can also be 
expressed in terms of $S_{\mu\nu}$ as
\begin{equation}
\nabla_{\nu}\hat{T}_{\mu}^{\nu}=-\nabla_{\nu}(S_{\mu\sigma}S^{\nu\sigma
})=-S_{\mu\sigma}\nabla_{\nu}S^{\nu\sigma}=0, \label{eqn:1036}
\end{equation}
where the uses have been made of $\nabla_{\nu}S_{\rho\sigma}+\nabla_{\rho
}S_{\sigma\nu}+\nabla_{\sigma}S_{\nu\rho}=0$ and $S_{\nu\rho}S^{\nu\rho}=0$
which is equivalent to (\ref{eqn:1027}) in Appendix A.
The corresponding quantity in Maxwell's EM theory $\nabla_{\nu}T_{\mu}^{\nu}=-F_{\mu\sigma}
\nabla_{\nu}F^{\nu\sigma}$ vanishes under the following condition of no electric current:
\begin{equation}
\nabla_{\nu}F^{\nu\sigma}=0,
\end{equation}
which yields the EM wave equation in the vacuum. Notice, however, that $\nabla_{\nu}S^{\nu\sigma}=0$ is
a sufficient condition for (\ref{eqn:1036}) but not a necessary one. According
to the argument in the previous section, a general form of $\nabla_{\nu}%
S^{\nu\sigma}$ that satisfies (\ref{eqn:1036}) is given by
\begin{equation}
\nabla_{\nu}S^{\nu\sigma}=aC^{\sigma}+Q^{\sigma}, \label{eqn:1038}
\end{equation}
where $Q^{\sigma}$ denotes a spacelike vector satisfying (\ref{eqn:1019}). 
Directly from the definition of $S_{\mu\nu}$, namely, the third equation in (\ref{eqn:1012}),
we have  
\begin{equation}
\nabla_{\nu}S_{\sigma}^{\nu}=-(\nabla_{\nu}L^{\nu})C_{\sigma}
+[C^{\nu}\nabla_{\nu}L_{\sigma}-L^{\nu}\nabla_{\nu}C_{\sigma}]. \label{eqn:2011}
\end{equation}
Since both $C_{\mu}$ and $L_{\mu}$ are gradient vectors satisfying the
integrability condition: $\nabla_{\nu}C_{\mu}=\nabla_{\mu}C_{\nu}$, the second
term in the square bracket is further rewritten as
\begin{eqnarray}
-L^{\nu}\nabla_{\nu}C_{\sigma} &  =-L^{\nu}\nabla_{\sigma}C_{\nu}
=-\nabla_{\sigma}(L^{\nu}C_{\nu})+C^{\nu}\nabla_{\sigma}L_{\nu}\nonumber\\
&  =C^{\nu}\nabla_{\nu}L_{\sigma},\nonumber
\end{eqnarray}
and hence, (\ref{eqn:2011}) becomes
\begin{equation}
\nabla_{\nu}S_{\sigma}^{\nu}=-(\nabla_{\nu}L^{\nu})C_{\sigma}+2C^{\nu}\nabla_{\nu}L_{\sigma}
=-(\nabla_{\nu}L^{\nu})C_{\sigma}, \label{eqn:1041}
\end{equation}
where the use has been made of (\ref{eqn:2001}). So, for CP flow under consideration, 
(\ref{eqn:1038}) reduces to
\begin{equation}
\nabla_{\nu}S^{\nu\sigma}= -(\nabla_{\nu}L^{\nu})C^{\sigma}. \label{eqn:1040}
\end{equation}
By similar manipulations, we also get 
\begin{eqnarray}
C^{\sigma}\nabla_{\sigma}S_{\mu\nu} &  =C^{\sigma}\nabla_{\sigma}(C_{\mu
}L_{\nu}-L_{\mu}C_{\nu})=C^{\sigma}C_{\mu}\nabla_{\sigma}L_{\nu}-C^{\sigma
}C_{\nu}\nabla_{\sigma}L_{\mu}\nonumber\\
&  + L_{\nu}C^{\sigma}\nabla_{\sigma}C_{\mu}-L_{\mu}C^{\sigma}\nabla_{\sigma}C_{\nu}=0.    
\label{eqn:1042}
\end{eqnarray}
Therefore we have the following important equation:
\begin{equation}
C^{\sigma}\nabla_{\sigma}S_{\mu\nu}=0. \label{eqn:1043}
\end{equation}
Note that the second equation in (\ref{eqn:1008}) and (\ref{eqn:1014}) are
rewritten, respectively, as 
\begin{equation}
C^{\nu}\nabla_{\nu}B=0;\,\,\,\,C^{\nu}\nabla_{\nu}\lambda=0. \label{eqn:1044}
\end{equation}
In the hydrodynamic terms, (\ref{eqn:1044}) tells us that, in addition to $B$ and $\lambda$, 
the vorticity $S_{\mu\nu}$ is also advected (or convected) along a null geodesic
with tangent vector $C^{\mu}$. Since the CP vortex dynamics is characterised by a couple of variables $B$ and $\lambda$,
(\ref{eqn:1043}) and (\ref{eqn:1044}) suggest that $S_{\mu\nu}$ can be parametrised in terms of co-moving 
Lagrange coordinates $B$ and $\lambda$ as
\begin{equation}
S_{\mu\nu}= S_{\mu\nu}(B, \lambda). \label{eqn:1045}
\end{equation}

Polarisation is an important aspect of EM waves which is related with spin dynamics, as
right and left circularly polarised states describe spin states of a photon.
To see how polarisation is represented in our formulation, 
we first note that the formulation on CP vortex dynamics can possess a dual parameter
space of $[(\lambda, B);(^{\star}\lambda, ^{\star}\phi (B))]$ illustrated in the following Fig. 1
where, without loss of generality, the spatial part of covector $C_{\mu}$ at the point of
interest is assumed
to be parallel to $x^{1}$ axis and that of covector $L_{\mu}$ lies on the plane spanned
by the two axes $x^{1}$ and $x^{3}$. In such a configuration, the magnetic counterpart
vector $\vec{M}_{(s)}$ in $S_{\mu\nu}$ system becomes parallel (or anti-parallel) to $x^{2}$ axis. The dual quantities corresponding to $C_{\mu}$ and $L_{\mu}$ together with
the associated definition of bivector and orthogonality condition are respectively
introduced as follows: 
\begin{eqnarray}
^{\star}L_{\mu} &  \equiv\nabla_{\mu}(^{\star}\lambda);\,\,\,\,^{\star}C_{\mu
}\equiv\nabla_{\mu}{^{\star}\phi}(B)=\varpi(B)C_{\mu},\label{eqn:2156}\\
^{\star}S_{\mu\nu} &  \equiv^{\star}C_{\mu}(^{\star}L_{\nu})-(^{\star}L_{\mu
})^{\star}C_{\nu},\,\,\,\,^{\star}L_{\nu}(^{\star}C^{\nu})=0,\label{eqn:3171}
\end{eqnarray}  
where $\varpi(B)\equiv d(^{\star}\phi)/d$B. The important points illustrated in 
Fig. 1 are that the spatial part of $^{\star}C_{\mu}$ is parallel (or anti-parallel)
to $C_{\mu}$ while $^{\star}L_{\mu}$ lies not on the plane spanned by $x^{1}$ and 
$x^{3}$ but on the one spanned by $x^{1}$ and $x^{2}$, from which ${^{\star}}\vec{M}_{(s)}$
becomes parallel (or anti-parallel) to $x^{3}$ axis.
A little bit lengthy but straightforward calculations using (\ref{eqn:1008}),
(\ref{eqn:1014}), (\ref{eqn:2156}) and (\ref{eqn:3171}) show that the orthogonality
condition holding between $\vec{M}_{(s)}$ and ${^{\star}}\vec{M}_{(s)}$ is equivalent to the
one between $L_{\mu}$ and $^{\star}L_{\mu}$, namely,
\begin{equation}
S_{23}(^{\star}S_{23})+S_{31}(^{\star}S_{31})+S_{12}(^{\star}S_{12}
)=-\varpi(B)(C_{0})^{2}L_{\nu}(^{\star}L^{\nu})=0.  \label{eqn:2172}
\end{equation}
\begin{figure}[!h]
\begin{center}
\unitlength 1mm
\begin{picture}(110, 65)
\put(40,35){\line(-1,-1){19}}
\put(40,35){\line(1,0){28}}
\put(40,35){\line(0,1){28}}
\thicklines
\put(40,35){\vector(-4,1){20}}
\put(40,35){\vector(-1,-1){11}}
\put(40,35){\vector(1,-1){13}}
\put(40,35){\vector(1,0){18}}
\put(40,35){\vector(0,1){18}}
\put(17,15){\makebox(0,0){$x^{1}$}}
\put(72,33){\makebox(0,0){$x^{2}$}}
\put(38,66){\makebox(0,0){$x^{3}$}}
\put(20,43){\makebox(0,0){$L_{\mu}$}}
\put(37,22){\makebox(0,0){$C_{\mu}\ (^{\star}C_{\mu})$}}
\put(57,22){\makebox(0,0){$^{\star}L_{\mu}$}}
\put(58,38){\makebox(0,0){$\vec{M_{(s)}}$}}
\put(46,53){\makebox(0,0){$\vec{^{\star}M_{(s)}}$}}
\put(5,5){Fig. 1: Dual configuration of $L_{\mu}$ and $^{*}L_{\mu}$ and associated}
\put(20, 0){ $\vec{M}_{(s)}$ and ${^{\star}}\vec{M}_{(s)}$}
\end{picture}
\end{center}
\label{fig:1}       
\end{figure}

Now, consider simple examples in a flat spacetime: a solution $S_{\mu\nu}$ of
the form:
\begin{equation}
\lambda=kx^{0}-kx^{1}+lx^{3};\,\,\,\,B=sin(kx^{0}-kx^{1}) \label{eqn:1048}
\end{equation}
where $k$ and $l$ are two positive constants. From (\ref{eqn:1048}), we get
the following non-zero components of $S_{\mu\nu}$:
\begin{equation}
E_{(s)}^{3}=S_{03}=klcos\theta_{k};\,\,\,\,M_{(s)}^{2}=S_{31}=klcos\theta_{k},
\label{eqn:1049}
\end{equation}
where $\theta_{k}\equiv k(x^{0}-x^{1})$. Clearly (\ref{eqn:1049}) is a
linearly polarised wave and the spatial part of covector $L_{\mu}$ lies in the
plane spanned by $C_{1}$ and $E_{(s)}^{3}$. In order to have a circularly
polarised wave, we need an additional linearly polarised one to superimpose on
it. This additional mode denoted by $(^{\star}\vec{E}_{(s)},^{\star}\vec{M}_{(s)})$ is
given by, say, 
\begin{equation}
^{\star}\lambda=kx^{0}-kx^{1}+lx^{2};\,\,\,\,^{\star}\phi=\sqrt{1-B^{2}
}=cos(kx^{0}-kx^{1})  \label{eqn:48}
\end{equation}
\begin{equation}
^{\star}E_{(s)}^{2}=^{\star}S_{02}=-klsin\theta_{k};\,\,\,^{\star}M_{(s)}
^{3}=^{\star}S_{12}=klsin\theta_{k},  \label{eqn:1050}
\end{equation}
where, as already pointed out, $^{\star}C_{\mu}$ is parallel (or
anti-parallel) to $C_{\mu}$ and the spatial part of $^{\star}L_{\mu}$ is now
not in the plane spanned by $C_{1}$ and $E_{(s)}^{3}$ but in the plane spanned
by $C_{1}$ and $M_{(s)}^{2}$. 
From (\ref{eqn:1049}) and (\ref{eqn:1050}), we see that $\vec{E}_{(s)}^{\pm}
\equiv \vec{E}_{(s)}\pm {^{\star}}\vec{E}_{(s)}; \,\,\vec{M}_{(s)}^{\pm}
\equiv \vec{M}_{(s)}\pm {^{\star}}\vec{M}_{(s)}$ correspond, respectively, to right (+) and left
(-) circular polarisations. According to quantum mechanics, a photon
corresponds either to right- or left-circularly polarised wave depending on
the sign of spin. In view of their direct relation with the spin degrees of freedom,
the combined states $(\vec{E}_{(s)}^{\pm};\,\,\vec{M}_{(s)}^{\pm})$ can be
taken as more fundamental than the linearly polarised states. A contrast
between circularly and linearly polarised waves can be seen in the absence for
the former modes of a nodal point where both of $\vec{E}_{(s)}$ and
$\vec{M}_{(s)}$ vanish at the same time. A circularly polarised wave has a
phase-independent constant amplitude of $|\vec{E}_{(s)}|=|\vec{M}_{(s)}|$, and
its \lq\lq Poynting\rq\rq  vector $\Pi_{(s)}^{\mu}$ defined in (\ref{eqn:1020})
and (\ref{eqn:1021}) also has a phase-independent
constant magnitude owing to
\begin{equation}
-\vec{E}_{(s)}^{\pm}\times\vec{M}_{(s)}^{\pm}=-\vec{E}_{(s)}\times\vec{M}_{(s)}
- {^{\star}}\vec{E}_{(s)}\times {^{\star}}\vec{M}_{(s)}=k^{2}l^{2},
\end{equation}
for the case given above. In (\ref{eqn:1034}), we see that the energy momentum tensor for $S_{\mu\nu}$
field is formally expressed as that of free moving particles in which $\rho$
denotes proper density or ``formal particle numbers in unit volume''. Since
$\rho = -L^{\nu}L_{\nu}=-g^{\mu\nu}\nabla_{\mu}\lambda\nabla_{\nu}\lambda$ 
corresponds to $l^{2}$ in the above example, the components of the
energy-momentum current per ``particle''and per wavelength becomes
proportional to $k$ in this hydrodynamic model, which can be regarded as the
classical version of Einstein - de Broglie relation in the $S_{\mu\nu}$
system. As clearly seen in the examples above, the four vectors $C^{\mu}$, $L^{\mu}$,
$^{\star}C^{\mu}$ and $^{\star}L^{\mu}$, satisfying the following mutually
orthogonal conditions (\ref{eqn:1051}) directly derived from (\ref{eqn:1014}),
 (\ref{eqn:2156}),(\ref{eqn:3171}) and (\ref{eqn:2172}), serve as the complete bases
 of $4d$ spacetime with which circularly polarised states can be constructed.
\begin{eqnarray}
C_{\nu}{^{\star}C^{\nu}}=0,\,\,\,\,\,L_{\nu}{^{\star}L^{\nu}}=0, \nonumber \\
C_{\nu}L^{\nu}=0,\,\,\,\,\,C_{\nu}{^{\star}L^{\nu}}=0,  \label{eqn:1051} \\
{^{\star}C_{\nu}}L^{\nu}=0,\,\,\,\,\,{^{\star}C_{\nu}}{^{\star}L^{\nu}}=0.
\nonumber
\end{eqnarray} 

It is worthwhile to point out that similar orthogonality conditions also hold
for the bivector components $S_{\mu\nu}$ if we consider them as the components
of dual spinor representations of $\Pi^{\mu}$ which can be derived from
(\ref{eqn:1034}). To see this, for simplicity, consider a linearly polarised
wave whose $\Pi^{\mu}$, $\vec{E}_{(s)}$ and $\vec{M}_{(s)}$ are oriented such 
that they are parallel (or anti-parallel), respectively, to the orthogonal
axes $x^{1}$, $x^{2}$ and $x^{3}$ of a Lorentz reference frame. (\ref{eqn:48})
and (\ref{eqn:1050}) serve as an example for this configuration. Since
$E_{(s)}^{2}=S_{02}$ and $M_{(s)}^{3}=S_{12}$, we have $\Pi^{1}=-S_{02}S_{12}%
$, using (\ref{eqn:1020}). Substituting this into the following definition of
spinor representation of $\Pi^{\mu}$ \cite{RefB2}, we obtain 
\begin{eqnarray}
\sqrt{2}\Psi(\Pi^{\mu})\equiv\left(
\begin{array}
[c]{cc}
\Pi^{0}+\Pi^{3} & \Pi^{1}+i\Pi^{2}\\
\Pi^{1}-i\Pi^{2} & \Pi^{0}-\Pi^{3}
\end{array}
\right)  =\left(
\begin{array}
[c]{cc}
\Pi^{0} & \Pi^{1}\\
\Pi^{1} & \Pi^{0}\label{eqn:1060}
\end{array}
\right)  .
\end{eqnarray}
As we mentioned in section 2, the reason why we confine ourself to real variables 
in the case of light-like modes is just for the sake of simplicity and full complexification
of variables is to be introduced for the discussion of non-light-like modes in the final section. 
So, we think that the usage of complex spinor representation like (\ref{eqn:1060}) in our
present discussion is not superficial though it may look so as far as we stay in the restricted representation by real variables.  
Using $(S_{02})^{2}=(S_{12})^{2}$, we get
\begin{eqnarray}
\sqrt{2}\Psi(\Pi^{\mu})=\left(
\begin{array}
[c]{cc}
(S_{02})^{2} & -S_{02}S_{12}\\
-S_{02}S_{12} & (S_{02})^{2}\label{eqn:1061}
\end{array}
\right)  ,
\end{eqnarray}
which becomes equal to the non-zero $2\times2$ minor matrix of degenerated
$\hat{T}^{\mu\nu}$: $\hat{T}_{(m)}^{MN};\,0\leq M,N\leq1$ (cf. (\ref{eqn:1034}
)). Equating (\ref{eqn:1061}) with $\hat{T}_{(m)}^{MN}$, we have
\begin{eqnarray}
\sqrt{2}\Psi(\Pi^{\mu})=\left(
\begin{array}
[c]{cc}
(S_{02})^{2} & -S_{02}S_{12}\\
-S_{02}S_{12} & (S_{12})^{2}
\end{array}
\right)  =\left(
\begin{array}
[c]{cc}
\sqrt{\rho}C^{0}\sqrt{\rho}C^{0} & \sqrt{\rho}C^{0}\sqrt{\rho}C^{1}\\
\sqrt{\rho}C^{1}\sqrt{\rho}C^{0} & \sqrt{\rho}C^{1}\sqrt{\rho}C^{1}  \label{eqn:1062}
\end{array}
\right). 
\end{eqnarray}
So, using (\ref{eqn:1034}), we see that the spinor representation of $\Pi^{\mu}$ 
also has a dual form in which no undetermined parameter is involved.
In appearance, the matrix components on the left-hand side of (\ref{eqn:1062})
look quite different from those on the right-hand side. In reality, however,
they are quite similar in the sense that both of them are represented as the
product of two vectors $C_{\mu}$ and $L_{\nu}$, since $\sqrt{\rho}$ is the
length of $L_{\nu}$, which justifies us to regard (\ref{eqn:1062}) as the unique dual
spionr representation of $\Pi^{\mu}$. 
So, we can say that ($\sqrt{\rho}C^{0}$, $\sqrt{\rho}C^{1}$,
$S_{02}$, $S_{12}$) is the complete orthogonal set of spinor components with which
not only complementary aspects of the radiation field can be described, but also
a set of orthogonal bases of spacetime is provided. 
In the subsequent section, we further show that $S_{\mu\nu}$ is not only endowed with the property
of the orthogonal bases of the spacetime but also with that of curvature of
the spacetime. Apparently, two vortex dynamics $F_{\mu\nu}$ and
$S_{\mu\nu}$ have different origins, nevertheless, as we have just shown here,
there exists a remarkable similarity between them, which implies that this
intrinsic geometric vortex mode provides a canonical form of the energy
propagation in the spacetime. 

\section{Properties of a vortex couplet}
\label{sec:3}

The Riemann curvature tensor satisfies the following properties: 
\begin{eqnarray}
R_{\mu\nu\rho\sigma}   =R_{[\mu\nu]\rho\sigma}=R_{\mu\nu\lbrack\rho\sigma
]};R_{\mu\nu\rho\sigma}=R_{\rho\sigma\mu\nu};\label{eqn:1751}\\
R_{\mu\lbrack\nu\rho\sigma]}   =0;\nabla_{\lbrack\mu}R_{\nu\rho]\sigma\tau
}=0.\label{eqn:1052}
\end{eqnarray}
The last two equalities in (\ref{eqn:1052}) are called the first and the
second Bianchi identities. If we introduce $\hat{S}_{\mu\nu\rho\sigma}%
=S_{\mu\nu}S_{\rho\sigma}$, then using (\ref{eqn:1017}), we readily see that
it satisfies:
\begin{eqnarray}
\hat{S}_{\mu\nu\rho\sigma}=\hat{S}_{[\mu\nu]\rho\sigma}=\hat{S}_{\mu\nu
\lbrack\rho\sigma]};\,\,\,\hat{S}_{\mu\nu\rho\sigma}=\hat{S}_{\rho\sigma\mu
\nu};\,\,\,\hat{S}_{\mu\lbrack\nu\rho\sigma]}=0.
\end{eqnarray}
As to the property corresponding to the second Bianchi identity, we first
calculate the quantity: $J_{\mu\nu\rho\sigma\tau}=\nabla_{\mu}\hat{S}_{\nu
\rho\sigma\tau}+\nabla_{\nu}\hat{S}_{\rho\mu\sigma\tau}+\nabla_{\rho}\hat
{S}_{\mu\nu\sigma\tau}$;
\begin{eqnarray}
J_{\mu\nu\rho\sigma\tau} &  =\nabla_{\mu}\hat{S}_{\nu\rho\sigma\tau}%
+\nabla_{\nu}\hat{S}_{\rho\mu\sigma\tau}+\nabla_{\rho}\hat{S}_{\mu\nu
\sigma\tau}\nonumber\label{eqn:1054}\\
&  =(\nabla_{\mu}S_{\nu\rho}+\nabla_{\nu}S_{\rho\mu}+\nabla_{\rho}S_{\mu\nu
})S_{\sigma\tau}+(S_{\nu\rho}\nabla_{\mu}+S_{\rho\mu}\nabla_{\nu}+S_{\mu\nu
}\nabla_{\rho})S_{\sigma\tau}\nonumber\\
&  =(S_{\nu\rho}\nabla_{\mu}+S_{\rho\mu}\nabla_{\nu}+S_{\mu\nu}\nabla_{\rho
})S_{\sigma\tau}
\end{eqnarray}
where the use has been made of an identity: $\nabla_{\mu}S_{\nu\rho}%
+\nabla_{\nu}S_{\rho\mu}+\nabla_{\rho}S_{\mu\nu}=0$. Secondly, referring to
(\ref{eqn:2202}), consider the Hodge dual of $S^{\mu\nu}$. In a local Lorentz
reference frame, its components $^{\ast}S_{(L)}^{\mu\nu}$ can be rewritten in
terms of those of $S_{\mu\nu}^{(L)}$. If we write down its components, we get
\begin{equation}
(^{\ast}S_{(L)}^{\mu\nu})=\left(
\begin{array}
[c]{cccc}
0 & S_{23}^{(L)} & S_{31}^{(L)} & S_{12}^{(L)}\\
-S_{23}^{(L)} & 0 & S_{03}^{(L)} & -S_{02}^{(L)}\\
\label{eqn:1055}-S_{31}^{(L)} & -S_{03}^{(L)} & 0 & S_{01}^{(L)}\\
-S_{12}^{(L)} & S_{02}^{(L)} & -S_{01}^{(L)} & 0
\end{array}
\right)  .
\end{equation}
By direct calculation, we obtain
\begin{equation}
(^{\ast}S_{(L)}^{\mu\nu}\partial_{\nu}\theta)=\left(
\begin{array}
[c]{c}
+(S_{23}^{(L)}\partial_{1}\theta+S_{31}^{(L)}\partial_{2}\theta+S_{12}%
^{(L)}\partial_{3}\theta)\\
-(S_{23}^{(L)}\partial_{0}\theta+S_{30}^{(L)}\partial_{2}\theta+S_{02}%
^{(L)}\partial_{3}\theta)\\
\label{eqn:1056}-(S_{31}^{(L)}\partial_{0}\theta+S_{10}^{(L)}\partial
_{3}\theta+S_{03}^{(L)}\partial_{1}\theta)\\
-(S_{12}^{(L)}\partial_{0}\theta+S_{20}^{(L)}\partial_{1}\theta+S_{01}%
^{(L)}\partial_{2}\theta)
\end{array}
\right)  ,
\end{equation}
whose every right-hand side component has the form of $\pm(S_{\nu\rho}%
^{(L)}\partial_{\mu}+S_{\rho\mu}^{(L)}\partial_{\nu}+S_{\mu\nu}^{(L)}%
\partial_{\rho})\theta$. Using the definition of $S_{\mu\nu}$ in
(\ref{eqn:1012}) and substituting $\lambda$ and $B$ into $\theta$ in the
above expression, we get
\begin{eqnarray}
(S_{\nu\rho}^{(L)}\partial_{\mu}+S_{\rho\mu}^{(L)}\partial_{\nu}+S_{\mu\nu
}^{(L)}\partial_{\rho})B=0;
(S_{\nu\rho}^{(L)}\partial_{\mu}+S_{\rho\mu}
^{(L)}\partial_{\nu}+S_{\mu\nu}^{(L)}\partial_{\rho})\lambda=0. \label{eqn:1057}
\end{eqnarray}
On the other hand, in the local Lorentz reference frame, (\ref{eqn:1054})
assumes the form
\begin{equation}
J_{\mu\nu\rho\sigma\tau}=(S_{\nu\rho}^{(L)}\partial_{\mu}+S_{\rho\mu}%
^{(L)}\partial_{\nu}+S_{\mu\nu}^{(L)}\partial_{\rho})S_{\sigma\tau}^{(L)}.
\label{eqn:1058}
\end{equation}
So substituting (\ref{eqn:1045}) into (\ref{eqn:1058})
and using (\ref{eqn:1057}), we finally get $J_{\mu\nu\rho\sigma\tau}=0$. Since
it is a tensor quantity, it must vanish in any other coordinate systems. Thus we get
\begin{equation}
\nabla_{\lbrack\mu}\hat{S}_{\nu\rho]\sigma\tau}=0.
\end{equation}

\section{Nakanishi-Lautrup formalism and on the coupling of EM and gravitational radiation}
\label{sec:4}

In this section, we discuss a possibility of synchronised energy
propagation of EM and gravitational fields on the basis of the results
obtained so far. As we will see shortly, the key ingredients of coupling the
energetics of two different fields are $C^{\mu}$ as a common component of the
two vortex dynamics and the Poynting vectors of the respective systems,
namely, $P^{\mu}$ and $\Pi^{\mu}$. In the particle-like representation,
$\Pi^{\mu}$ originally defined in (\ref{eqn:1020}) and (\ref{eqn:1021}) is
alternatively written as
\begin{equation}
\Pi^{\mu}=\rho C^{0}C^{\mu}.\label{eqn:1063}
\end{equation}
So far, the quantity $B$ introduced in (\ref{eqn:1008}) is purely geometrical
one. However, in EM theory, we can find a physical candidate for it. Consider
Maxwell equation in the vacuum: 
\begin{equation}
0=\nabla_{\nu}F^{\nu\rho}=-g^{\rho\sigma}\nabla_{\sigma}(\nabla_{\tau}A^{\tau
})+[g^{\sigma\tau}\nabla_{\sigma}\nabla_{\tau}A^{\rho}+R_{\sigma}^{\rho
}A^{\sigma}], \label{eqn:1064}
\end{equation}
which reduces to $0=[g^{\sigma\tau}\nabla_{\sigma}\nabla_{\tau}A^{\rho
}+R_{\sigma}^{\rho}A^{\sigma}]$ under the Lorentz gauge condition:
$\nabla_{\nu}A^{\nu}=0$. So, we assume that
\begin{equation}
g^{\sigma\tau}\nabla_{\sigma}\nabla_{\tau}A^{\rho}+R_{\sigma}^{\rho}A^{\sigma
}=0, \label{eqn:2501}
\end{equation}
which is one of the conventional forms of equation for $A^{\mu}$. What is not
conventional in the following discussion is the form of gauge condition which
is consistent with the energy-momentum conservation law. Referring to
(\ref{eqn:1036}), we get
\begin{equation}
\nabla_{\nu}{T}_{\mu}^{\nu}=-\nabla_{\nu}(F_{\mu\sigma}F^{\nu\sigma}%
)=-F_{\mu\sigma}\nabla_{\nu}F^{\nu\sigma}=0.
\end{equation}
So, we see that assuming (\ref{eqn:2501}) is equivalent to
\begin{equation}
-F_{\mu\sigma}\nabla_{\nu}F^{\nu\sigma}=F_{\mu\sigma}g^{\sigma\tau}%
\nabla_{\tau}B=0, \label{eqn:2503}
\end{equation}
where $B\equiv\nabla_{\rho}A^{\rho}$ should properly be identified with
Nakanishi's B-field \cite{RefJ2}. Since $\nabla_{\sigma}\nabla_{\nu}F^{\nu\sigma}$ vanishes
identically, we also have
\begin{equation}
g^{\sigma\tau}\nabla_{\sigma}\nabla_{\tau}B=0, \label{eqn:1066}
\end{equation}
which is equal to the first equation in (\ref{eqn:1008}). Comparing
[(\ref{eqn:2503}); (\ref{eqn:1066})] with [(\ref{eqn:1008}); (\ref{eqn:1015}
)], we see that 
\begin{equation}
B=\nabla_{\nu}A^{\nu}, \label{eqn:1166}
\end{equation}
is a mathematically consistent assumption which couples EM and $S_{\mu\nu}$
fields together through $A^{\mu}$.
Thus, we have shown that 
\begin{equation}
\nabla_{\nu}F^{\nu\sigma}+g^{\sigma\tau}\nabla_{\tau}B=0 \label{eqn:2505}
\end{equation} 
is not only mathematically consistent but physically relevant relation satisfying the energy-momentum
conservation law. 

A motivation of our study exploring the physical relevance of $A^{\mu}$ suggests us to regard 
 (\ref{eqn:2505}) not as an auxiliary mathematical constraint to remove redundant degree of freedom but as an excited physical
 state of $\nabla_{\nu}A^{\nu}$ whose
 \lq\lq ground state\rq\rq\,  is described by the non-divergent Lorentz gauge condition. 
In a conventional EM theory, any physically meaningful quantity is considered to be
directly tied with gauge invariance. 
In our new formulation, the same situation holds good if we only consider the restricted gauge transformation of the form:
\begin{equation}
\tilde{A}_{\mu} = A_{\mu} + \nabla_{\mu}\chi\,\, ;\,\,\,\,\, g^{\mu\nu}\nabla_{\mu}\nabla_{\nu} \chi =0,  \label{eqn:2506}
\end{equation} 
which is closely related to the conservation of energy-momentum tensor through (\ref{eqn:2503}) and (\ref{eqn:1066}). 
Directly from (\ref{eqn:1064}), we see that both of the first and the second terms on the r.h.s. are invariant under the above
gauge transformation.    
If we accept that $\nabla_{\nu}A^{\nu}$ is a physical quantity, then it must be treated on equal footing with $F^{\mu\nu}$, which is realised in the following 
Lagrangian approach. Consider a Lagrangian density of the form:
\begin{equation}
\mathfrak{L^{*}}=\mathfrak{L}+\mathfrak{L}_{GF}=-\frac{1}{4}F_{\mu\nu}F^{\mu\nu}-\frac{1}{2}(\nabla_{\nu}A^{\nu})^{2}, \label{eqn:2507}
\end{equation}  
then, from which variation with respect to $A_{\nu}$, we readily recover (\ref{eqn:2505}), namely,
\begin{equation}
(\nabla_{\mu}F^{\mu\nu}+g^{\nu\mu}\nabla_{\mu}B)\delta A_{\nu}=0. \label{eqn:2509}
\end{equation}
Although the second term in (\ref{eqn:2507}) is called as a gauge-fixing (GF) term while the first term is gauge-invariant, in our new formulation, both of them are \lq\lq gauge invariant\rq\rq\, in the sense
of (\ref{eqn:2506}). Interestingly enough, our classical way of introducing (\ref{eqn:2507}) based on the conservation of energy-momentum tensor is consistent with NL formalism for gauge field quantisation in which $\nabla_{\mu}A^{\mu}$ plays a key role for the equation of propagator to have meaningful solutions. 
In NL formalism, GF Lagrangian density is given by
\begin{equation}
\mathfrak{L}_{GF}= B\nabla_{\mu}A^{\mu}+\frac{\alpha}{2}B^{2}, \label{eqn:2511}
\end{equation}  
where $B$ field satisfies:
\begin{equation}
\nabla_{\mu}A^{\mu}+ \alpha B =0;\,\,\,\,\, g^{\sigma\tau}\nabla_{\sigma}\nabla_{\tau} B=0. \label{eqn:2513}
\end{equation}  
\rm By comparing (\ref{eqn:2507}) with [(\ref{eqn:2511}), (\ref{eqn:2513})], we get Feynman gauge of 
$\alpha =1$. As we have touched on in the introductory remarks, one should refer to Ojima \cite{RefJ3}
for macroscopic physical relevance of $\nabla_{\mu}A^{\mu}$. 

Based on the above arguments, we assume that the $B$ current defined as $C^{\sigma} \equiv g^{\sigma\tau}\nabla_{\tau}B$ is a physical entity at least macroscopically.  
Then, the dynamical system (\ref{eqn:1034}) is considered as a vortex geometrodynamics generated by physical $B$ current and
purely geometrical spacelike vector $L_{\mu}$. In the coupling between $F_{\mu\nu}$ and $S_{\mu\nu}$, $B$ current  
is perpendicular to both $F_{\mu\nu}$ and $S_{\mu\nu}$, which means that Poynting vector 
$P^{\mu}$ and its counterpart $\Pi^{\mu}$ are oriented to become parallel with each other.
So, corresponding to (\ref{eqn:1063}), there exists $\rho_{(EM)}$ which satisfies
\begin{equation}
P^{\mu}=\rho_{(EM)}C^{0}C^{\mu}.
\end{equation}
Thus for a given EM wave, $S_{\mu\nu}$ dynamics provides an additional energy-carrying freedom in the form of (\ref{eqn:1063}). 
The problem of gravitational radiations is usually considered within the framework of Einstein space for which Ricci 
tensor $R_{\mu\nu}$ vanishes. The immediate consequence of this assumption is the fact that, unlike EM
radiations, we cannot define an energy-momentum tensor for that radiation field. It is said that Einstein adopted the assumption $R_{\mu\nu}=0$ by 
referring to EM theory in which we have (\ref{eqn:1064}): $\nabla_{\nu}F^{\nu\rho}=0$.
However, a closer inspection given above shows that, since the actual spacetime as physical
vacuum is filled with ubiquitous EM radiation, we may have another possibility, namely, (\ref{eqn:2505}).
If we adopt it instead of $\nabla_{\nu}F^{\nu\rho}=0$, then, as an additional degree of freedom
in energy propagation, we have 
\begin{equation}
-R_{\mu\nu}=-S_{\mu\sigma}S_{\nu}^{\sigma}=\rho C_{\mu}C_{\nu}, \label{eqn:1169}
\end{equation}
which may be considered as gravitational radiation similar to that of EM field.
We note that there exists a similarity between the above coupling process through $B$-field and the unification of electric and magnetic fields in Maxwell theory through the introduction of electric displacement (ED) field 
since both of them are time-dependent quantities which are not only related to the conservation of the vector current $\nabla_{\nu}F^{\nu\sigma}$ but also are playing key roles in uniting different field.  
As is the case in EM theory, (\ref{eqn:1169}) for
steady states reduces to a familiar form of $R_{\mu\nu}=0$. Therefore, we
propose a hypothesis that gravitational radiation energy is carried by the
$S_{\mu\nu}$ field which is dual to $F_{\mu\nu}$. This hypothesis, among others,  does not require any substantial change in general relativity.
Since $\hat{S}_{\mu\nu\rho\sigma}$ defined as a vortex couplet behaves exactly like
Riemann curvature tensor, it may carry the elements of curvatures as well as
energy with the speed of light, which qualifies ($S_{\mu\nu}, F_{\mu\nu}$) 
as a candidate for energy-carrying gravito-electromagnetic (GEM) wave mode.
We conclude this section by pointing out the fact that the divergence of
Clebsch parametrised vector $\nabla_{\nu}(\lambda C^{\nu})$ vanishes under the
conditions of (\ref{eqn:1008}) and (\ref{eqn:1014}). Since the above procedure
of getting a dual structure ($F_{\mu\nu}$; $S_{\mu\nu}$) crucially depends
upon $\nabla_{\nu}A^{\nu}\neq0$, we cannot extend this procedure to get an
additional skew-symmetric field. 

\section{Vortex dynamics generated by time-like and space-like CP flows}
\label{sec:5}

The arguments so far developed are restricted to the dual vortex structure for light-like EM radiation
field whose energy-momentum tensor is given by (\ref{eqn:1033}) and that of the associated
$S_{\mu\nu}$ field is (\ref{eqn:1034}). In this section, we show that the newly introduced CP-flow formalism can be extended to the cases in which $U_{\mu}$ is either time-like or space-like. Using the vector symbols given in (\ref{eqn:1012}), modified CP flow vector 
originally defined in (\ref{eqn:1006}) becomes
\begin{equation}
U_{\mu}=\frac{1}{2}(\lambda C_{\mu}-\phi L_{\mu}), \label{eqn:1171}
\end{equation}   
where the symbol $B$ in (\ref{eqn:1012}) is now replaced by $\phi$.  As is the case for
radiation field, we assume that $U_{\mu}$ satisfies the geodesic equation (\ref{eqn:1003}),
which is written as
\begin{equation}
S_{\mu\nu}U^{\nu}+ \nabla_{\mu}V = 0,  \label{eqn:1173}
\end{equation}   
where $V \equiv U^{\nu}U_{\nu}/2$ and $S_{\mu\nu}$ remains to be the same as that given
in (\ref{eqn:1012}). In section 1, we referred to the normalisation of velocity four vector
with Lagrangian parametrisation (\ref{eqn:1005}) for which $V$ becomes either $0$ (for light-like velocity) or $\pm 1$ (for time-like/space-like one).  
In the case of light-like radiation field we have already discussed, the null condition of $U^{\nu}U_{\nu}=0$ is used naturally for both cases of Lagrangian and Clebsch parametrisations.
But the normalisation (\ref{eqn:1005}) which is self-evident in Lagrangian parametrisation
becomes moot in the case of CP flow. A natural normalisation condition within the framework of the present modified CP formalism would be attained through
the extension of (\ref{eqn:1008}) to the case of free non-light-like particle motions described
by the Klein-Gordon equation:
\begin{equation}
g^{\mu\nu}\nabla_{\mu}\nabla_{\nu}\psi + m^{2}\psi =0,  \label{eqn:1175}
\end{equation}   
where $m^{2}$ is a certain real scalar to be determined shortly.
A complex plane wave solution to (\ref{eqn:1175}) having the form of
$\psi= \exp{i(k_{\sigma}x^{\sigma})}$ satisfies the following a couple of equations:
\begin{equation}
g^{\mu\nu}\nabla_{\mu}\nabla_{\nu}\psi = -k^{\sigma}k_{\sigma}\psi;\;\;\;\;
g^{\mu\nu}\nabla_{\mu}\psi\nabla_{\nu}\psi = -k^{\sigma}k_{\sigma}\psi^{2}, \label{eqn:1177} 
\end{equation}
which is a natural extension of (\ref{eqn:1008}). So, here we assume that a couple of
(complex) Clebsch variables $\lambda$ and $\phi$ satisfy 
\begin{equation}
g^{\mu\nu}\nabla_{\mu}\nabla_{\nu}\psi = -m^{2}\psi;\;\;\;\;
g^{\mu\nu}\nabla_{\mu}\psi\nabla_{\nu}\psi = -m^{2}\psi^{2}, \label{eqn:1181} 
\end{equation}
and we see that the second equation
in (\ref{eqn:1181}) can be regarded as a sort of normalisation condition for covector $\nabla_{\mu}\psi$ 
in our modified CP flow formalism. For the plane wave solution mentioned above, since we have
\begin{equation}
g^{\mu\nu}\nabla_{\mu}\psi\nabla_{\nu}\psi^{*}=k^{\sigma}k_{\sigma}\psi\psi^{*},  \label{eqn:1182}
\end{equation} 
where $\psi^{*}$ denotes the complex conjugate of $\psi$, it is compatible to introduce
\begin{equation}
g^{\mu\nu}\nabla_{\mu}\psi\nabla_{\nu}\psi^{*}=m^{2}\psi\psi^{*},  \label{eqn:1184}
\end{equation} 
as the equation defining whether a given complex covector $\nabla_{\mu}\psi$ is time-like or not. 
Namely, since $\psi\psi^{*}$ is non negative, we can say that $\nabla_{\mu}\psi$ is time-like if $m^{2}>0$ and it is space-like if $m^{2}<0$, which corresponds to (\ref{eqn:1005}) for Lagrangian parametrisation of real vectors.
Substitution of $\lambda$ and $\phi$ into (\ref{eqn:1175}) yields
\begin{equation}
\nabla_{\nu}C^{\nu} + m^{2} \phi = 0;\;\;\;\;
\nabla_{\nu}L^{\nu} + m^{2} \lambda = 0, \label{eqn:1183} 
\end{equation}
and the second equation in (\ref{eqn:1181}) for $\phi$  and $\lambda$ respectively becomes
\begin{equation}
C^{\nu}C_{\nu} + m^{2} \phi^{2} = 0;\;\;\;\;
L^{\nu}L_{\nu} + m^{2} \lambda^{2} = 0. \label{eqn:1185} 
\end{equation}
Note that neither the first nor the second equations in (\ref{eqn:1181}) give a directional constraint on $C_{\mu}$ and $L_{\mu}$ so that, in addition to (\ref{eqn:1181}), we can impose on them the important orthogonality constraint already given in (\ref{eqn:1014}), but in our present
case, it must be defined in a complex form.
Defining $C^{\nu}=d^{\nu}+ie^{\nu}$ and $L_{\nu}=p_{\nu}+iq_{\nu}$, we get
\begin{equation}
C^{\nu}L_{\nu}= (d^{\nu}p_{\nu}-e^{\nu}q_{\nu})+i(d^{\nu}q_{\nu}+e^{\nu}p_{\nu})=0. \label{eqn:1186} 
\end{equation}
In $4d$ spacetime, it is always possible that we can specify the orientation of four vector $d^{\mu}$, $e^{\nu}$, $p^{\mu}$ and $q^{\mu}$ such that $d^{\nu}p_{\nu}=0$, $e^{\nu}q_{\nu}=0$, 
$d^{\nu}q_{\nu}=0$ and $e^{\nu}p_{\nu}=0$, which can be concisely rewritten as
\begin{equation}
C^{\nu}L_{\nu}= 0;\;\;\;\; C^{\nu}L_{\nu}^{*}=0, \label{eqn:1188} 
\end{equation}
where $L_{\nu}^{*}$ is the complex conjugate of $L_{\nu}$.
With this orthogonality condition, it can be readily shown that $U^{\nu}$ is a divergence free vector, namely, $\nabla_{\nu}U^{\nu}=0$.
Using (\ref{eqn:1185}) and (\ref{eqn:1188}), $V$ now becomes
\begin{equation}
V  =  (\frac{1}{2})^{3}(\lambda C^{\nu}-\phi L^{\nu})(\lambda C_{\nu}-\phi L_{\nu})
   =  -(\frac{1}{2})^{2}m^{2}(\lambda\phi)^{2}.   \label{eqn:1187}
\end{equation}

Now, going back to (\ref{eqn:1173}), direct calculations of $S_{\mu\nu}U^{\nu}$ and
$\nabla_{\mu}V$ yield
\begin{equation}
S_{\mu\nu}U^{\nu}+ \nabla_{\mu}V 
= -\frac{1}{4}(\lambda\phi)^{2}\nabla_{\mu}m^{2} =0,  \label{eqn:1189} 
\end{equation} 
which suggests that $m^{2}$ is not a variable but is a certain constant and, corresponding to (\ref{eqn:1005}),
we set $m^{2}=\pm m_{c}^{2}$  where $m_{c}$ is a real constant. 
By similar simple calculations, we also get 
\begin{equation}
U^{\sigma}\nabla_{\sigma}(\lambda\phi)=0;\;\;\;\; 
\Omega\equiv S_{\mu\nu}S^{\mu\nu}=2m^{4}(\lambda\phi)^{2},  \label{eqn:1191}
\end{equation}
from which we obtain an important advection equation:
\begin{equation}
U^{\sigma}\nabla_{\sigma}\Omega =0.  \label{eqn:1193}
\end{equation}
In section 3, we have looked into the form of energy-momentum tensor given by (\ref{eqn:1034})
based on (\ref{eqn:1033}). For non-lightlike case, if we follow the conventional EM knowledge again, 
it is natural to start with the form:
\begin{equation}
\hat{T}_{\mu}^{\nu} = -S_{\mu\sigma}S^{\nu\sigma} 
  + \frac{1}{4}S_{\alpha\beta}S^{\alpha\beta}g_{\mu}^{\nu}.  \label{eqn:1195}
\end{equation}
Through the well-known manipulation in EM theory, we get
\begin{equation}
\nabla_{\nu}\hat{T}_{\mu}^{\nu} = -S_{\mu\sigma}\nabla_{\nu}S^{\nu\sigma}. \label{eqn:1197}
\end{equation}
By quite similar manipulations deriving (\ref{eqn:2011}), we have
\begin{equation}
\nabla_{\nu}S^{\nu\sigma}=[-C^{\sigma}(\nabla_{\nu}L^{\nu})+L^{\sigma}(\nabla_{\nu}C^{\nu})]
+ [C^{\nu}\nabla_{\nu}L^{\sigma}-L^{\nu}\nabla_{\nu}C^{\sigma}].  \label{eqn:1199}
\end{equation}
In Appendix B, as in the case of deriving (\ref{eqn:1041}) from (\ref{eqn:2011})
using (\ref{eqn:1014}), we show that the contribution from the second term on the r.h.s. of 
(\ref{eqn:1199}) becomes naught, namely, 
\begin{equation}
-S_{\mu\sigma}[C^{\nu}\nabla_{\nu}L^{\sigma}-L^{\nu}\nabla_{\nu}C^{\sigma}]=0. \label{eqn:1201}
\end{equation}
Therefore, (\ref{eqn:1197}) becomes
\begin{eqnarray}
\nabla_{\nu}\hat{T}_{\mu}^{\nu} & = &
-S_{\mu\sigma}[-C^{\sigma}(\nabla_{\nu}L^{\nu})+L^{\sigma}(\nabla_{\nu}C^{\nu})] \nonumber \\
& = & -m^{2}S_{\mu\sigma}(\lambda C^{\sigma}-\phi L^{\sigma}) = -2m^{2}S_{\mu\sigma}U^{\sigma}.  \label{eqn:1203}
\end{eqnarray}
Using (\ref{eqn:1173}), (\ref{eqn:1187}) and the second equation in (\ref{eqn:1191}), the above leads to
\begin{equation}
\nabla_{\nu}\hat{T}_{\mu}^{\nu} = \nabla_{\nu}(-\frac{1}{4}\Omega g_{\mu}^{\nu}).  \label{eqn:1205}
\end{equation}
Combining (\ref{eqn:1195}) and (\ref{eqn:1205}) together with the notation $\hat{S}_{\mu\nu\sigma\rho}
=S_{\mu\nu}S_{\sigma\rho}$ used in section 5, we finally obtain
\begin{equation}
\nabla_{\nu}\hat{G}_{\mu}^{\nu}=0;\;\;\;\; 
  \hat{G}_{\mu}^{\nu} \equiv -\hat{S}_{\mu\sigma}^{\;\;\;\;\nu\sigma} 
  + \frac{1}{2}\hat{S}_{\alpha\beta}^{\;\;\;\;\alpha\beta}g_{\mu}^{\nu},  \label{eqn:1207}
\end{equation}
which is isomorphic to the Einstein equation: 
\begin{equation}
\nabla_{\nu}G_{\mu}^{\nu}=0;\;\;\;\; 
  G_{\mu}^{\nu} \equiv -R_{\mu\sigma}^{\;\;\;\;\nu\sigma} 
  + \frac{1}{2}R_{\alpha\beta}^{\;\;\;\;\alpha\beta}g_{\mu}^{\nu},  \label{eqn:1209}
\end{equation}
where $R_{\mu\nu\sigma\rho}$ denotes Riemann tensor.

The above geometrodynamic isomorphism suggests that there may exist a close link between vortex/spin
dynamics and spacetime structure. The notion suggesting such a link is not new and there has been quite
a few works inspired by Penrose's seminal paper \cite{RefJ5} on spin network. 
So in order to develop such an idea further, let us assume that the vortex model presented here is a
heuristic model of the spacetime and see what kind of information we can draw from it. By comparing
(\ref{eqn:1187}) and the second equation in (\ref{eqn:1191}), we have
\begin{equation}
\Omega = -8m^{2}V.  \label{eqn:1211}
\end{equation} 
Since our model based on the assumption (\ref{eqn:1181}) which admits complex variables, $\Omega$ and $V$ in the above equation are complex in general, so that we need a scheme for transforming
(\ref{eqn:1211}) into real physical variables. To do so, we first define $V_{(r)}$ as
\begin{equation}
V_{(r)} \equiv \frac{1}{2}U^{\nu}U_{\nu}^{*}
  =\frac{1}{8}(\lambda C^{\nu}-\phi L^{\nu})(\lambda^{*} C_{\nu}^{\nu}-\phi^{*} 
L_{\nu}^{*})=\frac{m^{2}}{4}(\lambda\lambda^{*})(\phi\phi^{*}),  \label{eqn:1213}
\end{equation}  
which is the proper measure for the magnitude of a given complex $U^{\mu}$ since $V_{(r)}$
becomes positive or negative depending on the sign of $m^{2}$ which respectively corresponds 
to time-like ($m^{2}=m_{c}^{2}$) and to space-like ($m^{2}=-m_{c}^{2}$) cases. The map: $V \longrightarrow V_{(r)}$
is obtained through multiplying $V$ by $-[(\lambda^{*}\phi^{*})/(\lambda\phi)]$ where we have
 minus sign because of the relation between the second equation in (\ref{eqn:1181}) and (\ref{eqn:1184}). Applying
 this map to (\ref{eqn:1211}), we have
\begin{equation}
\Omega_{(r)} \equiv  -2m^{4}(\lambda\lambda^{*})(\phi\phi^{*})
    = -8m^{2}V_{(r)}=-2m^{4}(\lambda\lambda^{*})(\phi\phi^{*})<0.  \label{eqn:1215}
\end{equation}     
The important point of (\ref{eqn:1215}) is that $\Omega_{(r)}$ is always negative regardless of
whether $U^{\mu}$ is space-like or not, which is consistent with the notion of negative vacuum energy called dark energy speculated in the context of accelerated expansion of spacetime\cite{RefJ6}. 
For (\ref{eqn:1193}), applying similar transformation from complex to real variables, we get 
\begin{equation}
U_{(r)}^{\nu} \equiv \frac{1}{2}[U^{\nu}+(U^{\nu})^{*}];\;\;\;
    U_{(r)}^{\nu}\nabla_{\nu}\Omega_{(r)}=0, \label{eqn:1217}
\end{equation}
where $U_{(r)}^{\nu}$ can either be time-like or space-like.

Finally, it should be pointed out that the modified CP flow formulation (\ref{eqn:1171})
valid either for time-like or for space-like $U_{\mu}$ can also cover a light-like case, if we change    
(\ref{eqn:1183}) and (\ref{eqn:1185}) such that
\begin{equation}
\nabla_{\nu}C^{\nu} \pm m_{c}^{2} \phi = 0;\;\;\;\;
\nabla_{\nu}L^{\nu} \mp m_{c}^{2} \lambda = 0, \label{eqn:1219} 
\end{equation}
and
\begin{equation}
C^{\nu}C_{\nu} \pm m_{c}^{2} \phi^{2} = 0;\;\;\;\;
L^{\nu}L_{\nu} \mp m_{c}^{2} \lambda^{2} = 0. \label{eqn:1221} 
\end{equation}
We can readily check the magnitude $V$ of $U_{\mu}$ given in (\ref{eqn:1187}) vanishes in
such a case and we see that  (\ref{eqn:1171}) reduces to a form given in  (\ref{eqn:1009}).
We note that (\ref{eqn:2003}) and (\ref{eqn:2007}) in the case of light-like modes correspond 
respectively to (\ref{eqn:1193}) and (\ref{eqn:1215}) in the non-light-like case, which shows
 that space-like quantities of $L^{\nu}L_{\nu}<0$ and $\Omega_{(r)}<0$ in both cases play a similar crucial 
role to impart energy to spacetime as the geometrical entity. 
\\
\\
\\
\noindent\bf Acknowledgements

\rm The author would like to express his sincere appreciation for valuable comments of Prof. Ojima at RIMS, Kyoto University, especially ones on B field in NL 
formalism and the interpretation of orthogonality condition:
$D=S_{01}S_{23}+S_{02}S_{31}+S_{03}S_{12}=0$ in terms of Pl\"{u}cker coordinates and for his long-term encouragement from the early stages of developing the key ideas on CP flows. 
\\
\\
\appendix
\renewcommand{\theequation}{A.\arabic{equation}}
\setcounter{equation}{0}
\noindent\textbf{Appendix A}\newline

Using (\ref{eqn:1018}), (\ref{eqn:1020}) and (\ref{eqn:1021}), we
readily have
\begin{equation}
(\Pi^{1})^{2}+(\Pi^{2})^{2}+(\Pi^{3})^{2}=S_{S}^{2}S_{T}^{2} \label{eqn:1022}
\end{equation}
where
\begin{equation}
S_{T}^{2} \equiv (S_{01})^{2}+(S_{02})^{2}+(S_{03})^{2}.
\end{equation}
With the uses of $\Pi^{\mu}$ and the following definition of helicity $h$:
\begin{equation}
h\equiv Q^{1}S_{23}+Q^{2}S_{31}+Q^{3}S_{12},
\end{equation}
(\ref{eqn:1019}) is rewritten as
\begin{eqnarray}
S_{01}Q^{1}+S_{02}Q^{2}+S_{03}Q^{3}  &  =0;\,\,\,\,-Q^{0}\Pi^{1}+\Pi^{0}
Q^{1}-hS_{23}=0\label{eqn:1025}\\
-Q^{0}\Pi^{2}+\Pi^{0}Q^{2}-hS_{31}  &  =0;\,\,\,\,-Q^{0}\Pi^{3}+\Pi^{0}
Q^{3}-hS_{12}=0. \label{eqn:1026}
\end{eqnarray}
Since $C^{\mu}$ is a solution to (\ref{eqn:1019}) and the helicity 
$h_{(C)}\equiv C^{1}S_{23}+C^{2}S_{31}+C^{3}S_{12}$ defined for it becomes zero,
from (\ref{eqn:1025}) and (\ref{eqn:1026}) we get
\begin{eqnarray}
S_{01}C^{1}+S_{02}C^{2}+S_{03}C^{3}  &  =0;\,\,\,\,-C^{0}\Pi^{1}+\Pi^{0}
C^{1}=0\label{eqn:2025}\\
-C^{0}\Pi^{2}+\Pi^{0}C^{2}  &  =0;\,\,\,\,-C^{0}\Pi^{3}+\Pi^{0}
C^{3}=0. \label{eqn:2026}
\end{eqnarray}
So, we have
\begin{equation}
(-C^{0}\Pi^{1}+\Pi^{0}C^{1})^{2}+(-C^{0}\Pi^{2}+\Pi^{0}C^{2})^{2}+(-C^{0}\Pi^{3}+\Pi^{0}C^{3})^{2}=0.
\label{eqn:2027}
\end{equation}
Using (\ref{eqn:1022}), the first equality in (\ref{eqn:1020}), the repeated use of (\ref{eqn:2025}) and 
(\ref{eqn:2026}), (\ref{eqn:2027}) is further rewritten as 
$S_{S}^{2}S_{T}^{2}(C^{0})^{2}=S_{S}
^{4}[(C^{1})^{2}+(C^{2})^{2}+(C^{3})^{2}]$, from which we get
\begin{equation}
S_{T}^{2}=S_{S}^{2}, \label{eqn:1027}
\end{equation}
since $C^{\mu}$ is a null vector. A couple of conditions of (\ref{eqn:1018})
and (\ref{eqn:1027}) are exactly the same as those of EM radiation field: $\vec{E}%
\perp\vec{B}$ and $\parallel\vec{E}\parallel=\parallel\vec{B}\parallel$. By
using (\ref{eqn:1027}), the first equation in (\ref{eqn:1020}) is rewritten in the same
form as the energy density of EM field:
\begin{equation}
\Pi^{0}=\frac{1}{2}[(S_{01})^{2}+(S_{02})^{2}+(S_{03})^{2}+(S_{23}
)^{2}+(S_{31})^{2}+(S_{12})^{2}],
\end{equation}
and again from the first equation in (\ref{eqn:1020}), (\ref{eqn:1022}) and
(\ref{eqn:1027}), we readily see that vector $\Pi^{\mu}$ is a null vector.
Furthermore, since $h_{(\Pi)}\equiv\Pi^{1}S_{23}+\Pi^{2}S_{31}+\Pi^{3}
S_{12}=0$, we also see from the second equation in  (\ref{eqn:1025}) and
(\ref{eqn:1026}) that $\Pi^{\mu}$ satisfies (\ref{eqn:1019}). Thus we have
shown that $\Pi^{\mu}$ is parallel to $C^{\mu}$. 
Repeating the same procedures directly applied to $Q^{\mu}$ without replacing $Q^{\mu}$
by $C^{\mu}$ in the above derivations beginning from (\ref{eqn:2025}) and leading to (\ref{eqn:1027}),
 we obtain
\begin{equation}
\Pi^{0}[(Q^{0})^{2}-(Q^{1})^{2}-(Q^{2})^{2}-(Q^{3})^{2}]=-h^{2},
\label{eqn:1029}
\end{equation}
which shows that the vector $Q^{\mu}$ with non-zero $h$ is space-like.
\\
\\

\renewcommand{\theequation}{B.\arabic{equation}}
\setcounter{equation}{0}
\noindent\textbf{Appendix B}\newline

Here we show the equation
\begin{equation}
I=-S_{\mu\sigma}(C^{\nu}\nabla_{\nu}L^{\sigma}-L^{\nu}\nabla_{\nu}C^{\sigma})=0.  \label{eqn:1301}
\end{equation}
Using the definition of $S_{\mu\sigma}$ given in (\ref{eqn:1012}), we have
\begin{eqnarray}
I & = & (L_{\mu}C_{\sigma}-L_{\sigma}C_{\mu})(C^{\nu}\nabla_{\nu}L^{\sigma}-L^{\nu}\nabla_{\nu}C^{\sigma}) 
   \nonumber \\
  & = & L_{\mu}C_{\sigma}C^{\nu}\nabla_{\nu}L^{\sigma}-L_{\mu}C_{\sigma}L^{\nu}\nabla_{\nu}C^{\sigma}
  -L_{\sigma}C_{\mu}C^{\nu}\nabla_{\nu}L^{\sigma}  \nonumber \\
  & + & L_{\sigma}C_{\mu}L^{\nu}\nabla_{\nu}C^{\sigma} \nonumber \\
  & = & -\frac{1}{2}L_{\mu}L^{\nu}\nabla_{\nu}(C^{\sigma}C_{\sigma})
        -\frac{1}{2}C_{\mu}C^{\nu}\nabla_{\nu}(L^{\sigma}L_{\sigma})  \nonumber \\
  & + & L_{\mu}C_{\sigma}C^{\nu}\nabla_{\nu}L^{\sigma} + L_{\sigma}C_{\mu}L^{\nu}\nabla_{\nu}C^{\sigma}.
 \label{eqn:1303}
\end{eqnarray}
Utilising the integrability condition explained just after (\ref{eqn:2011}), the third and fourth terms in (\ref{eqn:1303}) are further rewritten as
\begin{eqnarray}
L_{\mu}C_{\sigma}C^{\nu}\nabla_{\nu}L^{\sigma} & = & L_{\mu}C^{\nu}[\nabla_{\nu}(C_{\sigma}L^{\sigma})
  -L^{\sigma}\nabla_{\nu}C_{\sigma}] \nonumber \\
  & = & - L_{\mu}C^{\nu}L^{\sigma}\nabla_{\nu}C_{\sigma}= -L_{\mu}C^{\nu}L^{\sigma}\nabla_{\sigma}C_{\nu}
  \nonumber \\
  & = & - \frac{1}{2}L_{\mu}L^{\sigma}\nabla_{\sigma}(C^{\nu}C_{\nu}).   \label{eqn:1305}
\end{eqnarray}
\begin{eqnarray}
L_{\sigma}C_{\mu}L^{\nu}\nabla_{\nu}C^{\sigma} & = & C_{\mu}L^{\nu}[\nabla_{\nu}(L_{\sigma}C^{\sigma}) 
  -C^{\sigma}\nabla_{\nu}L_{\sigma}] \nonumber \\
  & = & - C_{\mu}L^{\nu}C^{\sigma}\nabla_{\nu}L_{\sigma}= -C_{\mu}L^{\nu}C^{\sigma}\nabla_{\sigma}L_{\nu}
  \nonumber \\
  & = & - \frac{1}{2}C_{\mu}C^{\sigma}\nabla_{\sigma}(L^{\nu}L_{\nu}).   \label{eqn:1307}
\end{eqnarray}
Substituting (\ref{eqn:1305}) and (\ref{eqn:1307}) into (\ref{eqn:1303}), we have
\begin{eqnarray}
I & = & -L_{\mu}L^{\nu}\nabla_{\nu}(C^{\sigma}C_{\sigma})-C_{\mu}C^{\nu}\nabla_{\nu}(L^{\sigma}L_{\sigma})
  \nonumber \\
  & = & L_{\mu}L^{\nu}\nabla_{\nu}(m\phi^{2}) + C_{\mu}C^{\nu}\nabla_{\nu}(m\lambda^{2}) \nonumber \\
  & = & 2m(\phi L_{\mu} + \lambda C_{\mu})(L^{\nu}C_{\nu}) = 0,   \label{eqn:1309}
\end{eqnarray}
where the uses have been made of (\ref{eqn:1185}) and (\ref{eqn:1014}).


\begin{thebibliography}{}

\bibitem{RefJ1}
Tonomura, A., Osakabe, N., Matsuda, T., Kawasaki, T., Endo, J., Yano, S. and
Yamada, H. : Evidence for Aharanov-Bohm effect with magnetic field completely
shielded from electron wave. Phys. Rev. Lett., 56, 792-795, (1986)\\

\bibitem{RefJ2}
Nakanishi, N. : Prog. Theor. Phys. 35. 1111 (1966); Prog. Theor. Phys. 38. 881, (1967) \\

\bibitem{RefJ3}
Ojima, I. : Nakanishi-Lautrup $B$-filed, crossed product and duality,
RIMS workshop \lq\lq Research on Quantum Field Theory\rq\rq\, (2006)\\

\bibitem{RefJ4}
Penrose, R. : Twistor Algebra, J. Math. Physics 8, 345, (1967) \\

\bibitem{RefB1}
Penrose, R. and Rindler, W.: Spinors and space-time, vols. 1 and 2, Cambridge Univ. Press, (1984)\\

\bibitem{RefB2}
Huggett, S. A. and Tod, K. P.: An introduction to Twistor theory 2nd Ed., pp. 9.
Cambridge Univ. Press, (1994)\\


\bibitem{RefB3}
Lamb, H.: Hydrodynamics 6th Ed., pp. 248-249. Cambridge Univ. Press, (1932)\\

\bibitem{RefJ7}
Ojima, I. : In private communications.\\



 


\bibitem{RefJ5}
Penrose, R. : Angular momentum: an approach to combinatorial space-time, in Quantum Theory
and Beyond. Cambridge Univ. Press, (1971) \\

\bibitem{RefJ6}
Goldhaber, G and Perimutter, S: Astudy of 42 type la supernovae and a resulting measurement
of Omega(M) and Omega(Lambda),Physics Reports-Review section of Physics Letters
307 (1-4): 325-33, (1998)  \\







\end{thebibliography}
\end{document}